\documentclass{emulateapj}
\usepackage{natbib}
\def\thesource{{3C~454.3}}
\def\feii{{Fe~{\sc ii}}}
\def\mgii{{Mg~{\sc ii}}}
\def\feii{{Fe~{\sc ii}}}
\def\halpha{\ifmmode {\rm H}\alpha \else H$\alpha$\fi}
\def\civ{\ifmmode {\rm C{\sc iv}} \else C~{\sc iv}\fi}
\def\gsim{\lower 2pt \hbox{$\, \buildrel {\scriptstyle >}\over
{\scriptstyle \sim}\,$}}
\def\lsim{\lower 2pt \hbox{$\, \buildrel {\scriptstyle <}\over
{\scriptstyle \sim}\,$}}

\slugcomment{Submitted to ApJ Letters}

\shortauthors{Bonning et al.}
\shorttitle{Correlated Variability in 3C~454.3}
\begin{document}

\title{Correlated Variability in the Blazar 3C~454.3}

\author{E.~W. Bonning\altaffilmark{1}, 
C. Bailyn\altaffilmark{2}, 
C. M. Urry\altaffilmark{1}, 
M. Buxton\altaffilmark{2}, 
G. Fossati\altaffilmark{4}, 
L. Maraschi\altaffilmark{3}, 
P. Coppi\altaffilmark{2}, 
R. Scalzo\altaffilmark{1},
J. Isler\altaffilmark{1},
A. Kaptur\altaffilmark{2}}

\altaffiltext{1}{Department of Physics and Yale Center for Astronomy
and Astrophysics, Yale University, PO Box 208121, New Haven, CT 06520-8121;
erin.bonning@yale.edu}
\altaffiltext{2}{Department of Astronomy and Yale Center for Astronomy
and Astrophysics, Yale University, PO Box 208101, New Haven, CT 06520-8101}
\altaffiltext{3}{INAF - Osservatorio Astronomico di Brera, V. Brera 28, I-20100 Milano, Italy}
\altaffiltext{4}{Department of Physics and Astronomy, Rice University, Houston, TX 77005}

\begin{abstract}
The blazar 3C~454.3 was revealed by the Fermi Gamma-ray Space
Telescope to be in an exceptionally high flux state in July 2008.
Accordingly, we performed a multi-wavelength monitoring campaign on
this blazar using IR and optical observations from the SMARTS
telescopes, optical, UV and X-ray data from the Swift satellite,  
and public-release gamma-ray data from Fermi. We find an excellent
correlation between the IR, optical, UV and gamma-ray light curves, 
with a time lag of less than one day. The amplitude of the infrared
variability is comparable to that in gamma-rays, 
and larger than at optical or UV wavelengths.
The X-ray flux is not strongly correlated with either the gamma-rays
or longer wavelength data.  These variability characteristics find a
natural explanation in the external Compton model, in which electrons
with Lorentz factor  $\gamma \sim 10^{3-4}$ radiate synchrotron
emission in the infrared-optical and also scatter accretion disk or
emission line photons  to gamma-ray energies, while much cooler
electrons ($\gamma \sim 10^{1-2}$) produce X-rays by scattering
synchrotron or other ambient photons.  
\end{abstract}


\section{Introduction}

\label{sec:intro} 
Blazars are understood to be active galactic nuclei (AGN)
with aligned relativistic jets \citep{urry95}, so they offer a unique laboratory
for studying the physics of astrophysical jets. 
The spectral energy distributions (SEDs) of blazars
have a characteristic double-humped shape with a low-energy component
peaking anywhere from radio to X-rays, and a high-energy component
peaking at MeV to GeV energies \citep{fossati98}.
Flat Spectrum Radio Quasars (FSRQs) like 3C~454.3 
have SED peaks at radio-IR wavelengths and $\sim 1$~GeV \citep{urry95,
sambruna96}. The low-energy component is well
modeled as synchrotron emission from relativistic electrons in the jet
\citep{konigl81, urry82}, while the origin of the second SED peak at high
energies is not fully understood. Current explanations for the gamma-ray emission fall into two categories, leptonic and hadronic.
Leptonic models produce high-energy flux by inverse-Compton scattering of low-energy seed photons, 
either the synchrotron photons themselves \citep[Synchrotron
Self-Compton, ][]{jones74} 
or photons from an external source, such as thermal accretion disk emission or broad-line emission 
\citep{sikora94, dermer93, ghisellini96, celotti08}. In hadronic models,
protons that are accelerated to very high energies in the jet
produce gamma-rays from neutral pion decay, proton
synchrotron emission, and synchrotron emission from pair production
\citep{mucke01, mucke03, boettcher07}. Both leptonic and hadronic models can adequately fit 
single-epoch blazar SEDs, but variability offers a test of either model.

\thesource\ was among the more intense and variable FSRQs
detected with CGRO EGRET \citep{hartman99}, varying over
several years by factors of up to five, with a flare-state flux of
$F_{\rm >100~MeV} \sim$0.5$\times$10$^{-6}$~photons/s/cm$^2$
\citep{hartman93,hartman99,aller97}.
Long-term optical variability has also been reported, with up to
$\sim$3 mag changes over several years \citet{djorgovski08}.
During a 2005 optical flare to R=12 \citep{villata06},
\thesource\ was detected with INTEGRAL at a flux of 
$F_{\rm 3-200~keV} \sim$3$\times$10$^{-2}$~photons/s/cm$^2$ \citep{pian06}; 
a radio flare followed about a year later \citep{villata07}. 
\thesource\ has been detected with the AGILE gamma-ray satellite 
\citep{tavani08}, flaring in July 2007 and again in July 2008 
\citep{vercellone08,gasparrini08} with associated flaring at optical
and longer wavelengths \citep{ghisellini07,villata08}. On 24 July
2008, \citet{tosti08} confirmed the high gamma-ray flux state of the
source with a detection by the Fermi Large Area 
Telescope (LAT) while still in its post-launch commissioning phase. 
In the Fermi/LAT first light image released on 26 August 2008,
\thesource\ was among the brightest sources in the gamma-ray sky,
at the high end of its recorded gamma-ray intensity,
$F_{\rm 0.1-300~GeV} \sim$4.4$\times$10$^{-6}$ photons/s/cm$^2$.

Here we present data from our multi-wavelength optical and infrared
monitoring program of \thesource\ from June to December 2008 with the
Small and Moderate Aperture Research Telescope System (SMARTS). 
We correlate these data with Target of Opportunity observations 
carried out with the Swift X-ray Telescope (XRT) and 
Ultraviolet and Optical Telescope (UVOT), as
well as with 0.1--300 GeV fluxes made public by the Fermi Science
Support Center. The observations are described in Section~\ref{sec:obs}. The light curves, 
correlation functions, and SED are discussed in Section~\ref{sec:results}.

\section{Observations}
\label{sec:obs}

\subsection{SMARTS}
\label{sec:smarts}

Photometric monitoring of \thesource\ was carried out on the 1.3m
telescope located at Cerro Tololo Interamerican Observatory (CTIO)
with the ANDICAM instrument. ANDICAM is a dual-channel imager with a
dichroic that feeds an optical CCD and an IR imager, which can obtain
simultaneous data from 0.4 to 2.2 $\mu$. Our campaign began with
observations in B, V, R and J-bands with a cadence of one observation every 2 nights. 
After it became clear that 3C~454.3 was exhibiting
interesting and varied behavior, we added K-band observations
and increased the cadence to one observation every night. The SMARTS
photometric data and light curves for \thesource\ as well as all other
Fermi/LAT monitored blazars visible from CTIO are made publicly
available on a 1-2 day timescale on the web. \footnote{http://astro.yale.edu/glast/index.html}

Optical data were bias-subtracted, overscan-subtracted, and flat
fielded using {\it ccdproc} in IRAF. The optical photometry was calibrated
using published magnitudes of a secondary standard star\footnote{Shown
as star H in the finding chart at
http://www.lsw.uni-heidelberg.de/projects/extragalactic/charts/2251+158.html}
in the field of 3C~454.3 \citep{craine71, angione71, fiorucci98}.
Infrared data were sky-subtracted, flat fielded, and dithered images
combined using in-house IRAF scripts. The infrared photometry was
calibrated using 2MASS magnitudes of a secondary standard star (the
same star used in optical photometry calibration) in the field of 
3C~454.3. We estimated photometric errors by calculating the 1-$\sigma$
variation in magnitude of comparison stars with comparable magnitude
to 3C~454.3. These are as follows: B$_{\rm err}$~=~0.02 mag, V$_{\rm
err}$~=~0.02 mag, R$_{\rm err}$~=~0.02, J$_{\rm err}$~=~0.04 mag, and
K$_{\rm err}$~=~0.04 mag.

Figure~\ref{fig:lc} shows the B-band light curve normalized to its
flux at JD~2454700. Figure~\ref{fig:sed} shows two SEDs for
\thesource: one averaged over the actively flaring period up to
JD~2454750, and a second averaged over the relatively quiescent period
after that day. To compute the fluxes,  magnitudes were dereddened
using the extinction relations in \citet{cardelli89} together
with the value for A$_{\rm B}$ given by \citet{schlegel98} and
converted into flux densities using the zero-point fluxes given by
\citet{bessell98} and \citet{beckwith76}

\subsection{Fermi}
\label{sec:fermi}

The Fermi Space Telescope (formerly GLAST) was launched on 11 June
2008. The Fermi observatory Large Area Telescope (LAT) is designed to
measure the cosmic gamma-ray flux up to $\sim$ 300 GeV. The LAT is an
imaging, wide field-of-view high-energy pair conversion telescope
with energy range from $\sim$ 20 MeV to $\gsim$ 300 GeV, and surveys
the sky every three hours \citep{michelson07}. As a service to the
community and in order to support correlated multiwavelength
observations, the LAT Instrument Science Operations Center provides daily and
weekly averaged fluxes for a number of blazars, of which \thesource\ is
one.
Fluxes and 1$\sigma$ uncertainties for three bands, 0.1--300 GeV, 0.3--1 GeV, and 1--300 GeV, 
using preliminary instrument response functions and calibrations, are
made available online roughly once per week, with the caveat that the
early flux estimates are not absolutely calibrated, and may have
variations of up to 10\% due to uncorrected systematic effects. 
Because the observed variations are well correlated
with independently measured IR, optical, and UV variations,
we conclude the gamma-ray variations will not change significantly
even if they are eventually recalibrated, and in any case, 
our key results are robust against 10\% fluctuations in gamma-ray
intensity. We show the \thesource\ light curve in the 0.1--300 GeV
band in Figure~\ref{fig:lc} normalized to its photon flux at
JD~2454700. Fluxes shown in Figure~\ref{fig:sed} are computed from the
publicly released data in the 0.3--1 GeV and 1--300 GeV bands by
assuming a power-law spectrum of photon index $\Gamma$=2.

\subsection{Swift}
\label{sec:swift}

Since being identified in June 2008 as an extraordinarily bright
gamma-ray source \citep{vittorini08, gasparrini08},
\thesource\ has been the subject of numerous Swift target of opportunity
observations, including one by PI Bonning covering 22 September -
02 October, 2008. 
The Swift satellite \citep{gehrels04} has three instruments: 
a coded-mask Burst Alert Telescope \citep[BAT,][]{barthelmy05},
an X-ray Telescope covering the energy range 0.2--20 keV \citep[XRT,][]{burrows05},
and an Ultraviolet/Optical Telescope covering 170--600 nm \citep[UVOT,][]{roming05}.
Swift data are made public to the community
within a few days of the observations; therefore we were able to collect all
available data within the period of our SMARTS observations. We
reduced the data from the X-ray telescope (XRT) and the Ultraviolet
Optical Telescope (UVOT) according to the standard recipes given by
the Swift data analysis manuals.

For each obsid, the UVOT data for each exposure were co-added with the
task {\it uvotimsum}. The source magnitudes were then computed from a
source region of 5.5 arcsec using the task {\it uvotsource}, which
performs aperture photometry on the source and returns the count rate, flux density, and
magnitude in the Swift/UVOT photometric system \citep{poole08}.  We
correct these for interstellar extinction as described in
Section~\ref{sec:smarts}. 
Light curves from the UVOT B and W1 bands are shown in
Figure~\ref{fig:lc}, and average fluxes before and after JD~2454750 in
Figure~\ref{fig:sed}.

For each obsid, the XRT level-2 event list was generated via {\it
xrtpipeline v. 0.11.5} with the default filtering and screening
criteria, selecting photon counting (PC) data with XRT event grades
0-12. We extracted the source spectrum from a region centered at the
source with a radius of 60 arcsec and subtracted the background from a
nearby source-free region. Spectra were rebinned to 25 cts/bin, fit
with an absorbed power law, and the flux was computed 
in $0.5-2.0$ keV and $2.0-10.0$ keV bands. 
The X-ray light curve is shown in the bottom panel of Figure~\ref{fig:lc}.

\section{Results and Discussion}
\label{sec:results}

\begin{figure}[]
\begin{center}
\plotone{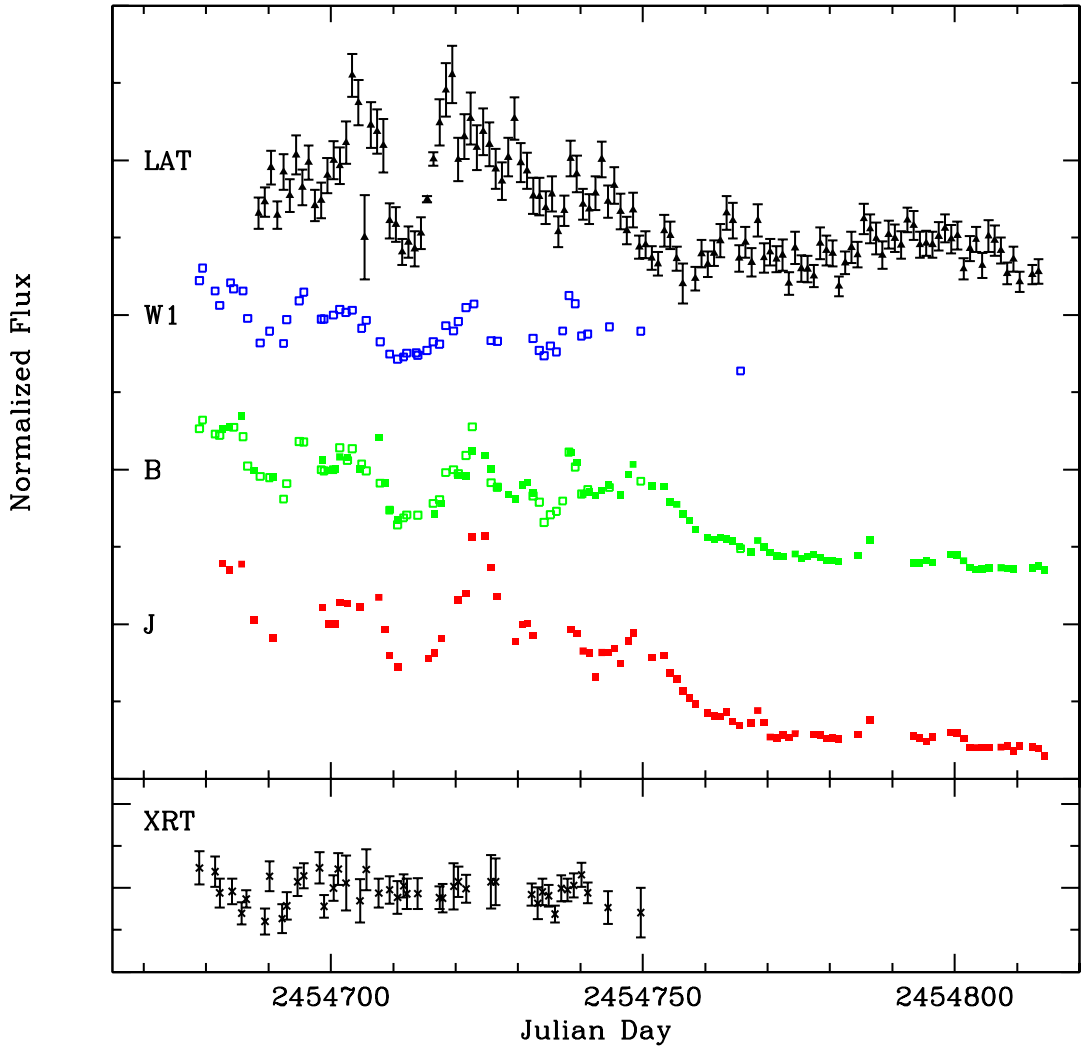} 
\figcaption[fig1]{Multi-wavelength light curves of \thesource\  
at ({\it top panel}) gamma-ray (0.1--300~GeV), UV (W1), optical (B), 
and IR (J) wavelengths from Fermi LAT, Swift UVOT, and SMARTS.
Fluxes have been normalized to JD~2454700. Light curves are offset for
clarity; minor tick spacing corresponds to 50\% change. 
Fluxes at JD~2454700 are 
2.83$\times$10$^{-6}$~cts~s$^{-1}$ at 0.1--300 GeV,
1.64$\times$10$^{-11}$~erg~s$^{-1}$~cm$^{-2}$ in W1,
2.21$\times$10$^{-11}$~erg~s$^{-1}$~cm$^{-2}$ in B, and
3.62$\times$10$^{-11}$~erg~s$^{-1}$~cm$^{-2}$ in J. 
({\it Bottom panel}) Swift XRT 2-10 keV light curve, normalized to flux at
JD~2454700 (2.90$\times$10$^{-11}$~erg~s$^{-1}$~cm$^{-2}$). 
The IR/optical/UV variations are well correlated with the 
gamma-ray variations, with a lag of $\lesssim 1$~day,
while the (minimal) X-ray variability is uncorrelated.
The variability has much higher amplitude in the J-band
than in B, which can be explained if there is an
relatively constant blue component, as expected for an accretion disk.
At z=0.859, Balmer continuum from an accretion disk, as well as
\feii\ and \mgii\ emission lines would be redshifted into the B and V
bands; H$\alpha$ is shifted into the J band. 
}
\label{fig:lc}
\end{center}
\end{figure}

\begin{figure}[]
\begin{center}
\plotone{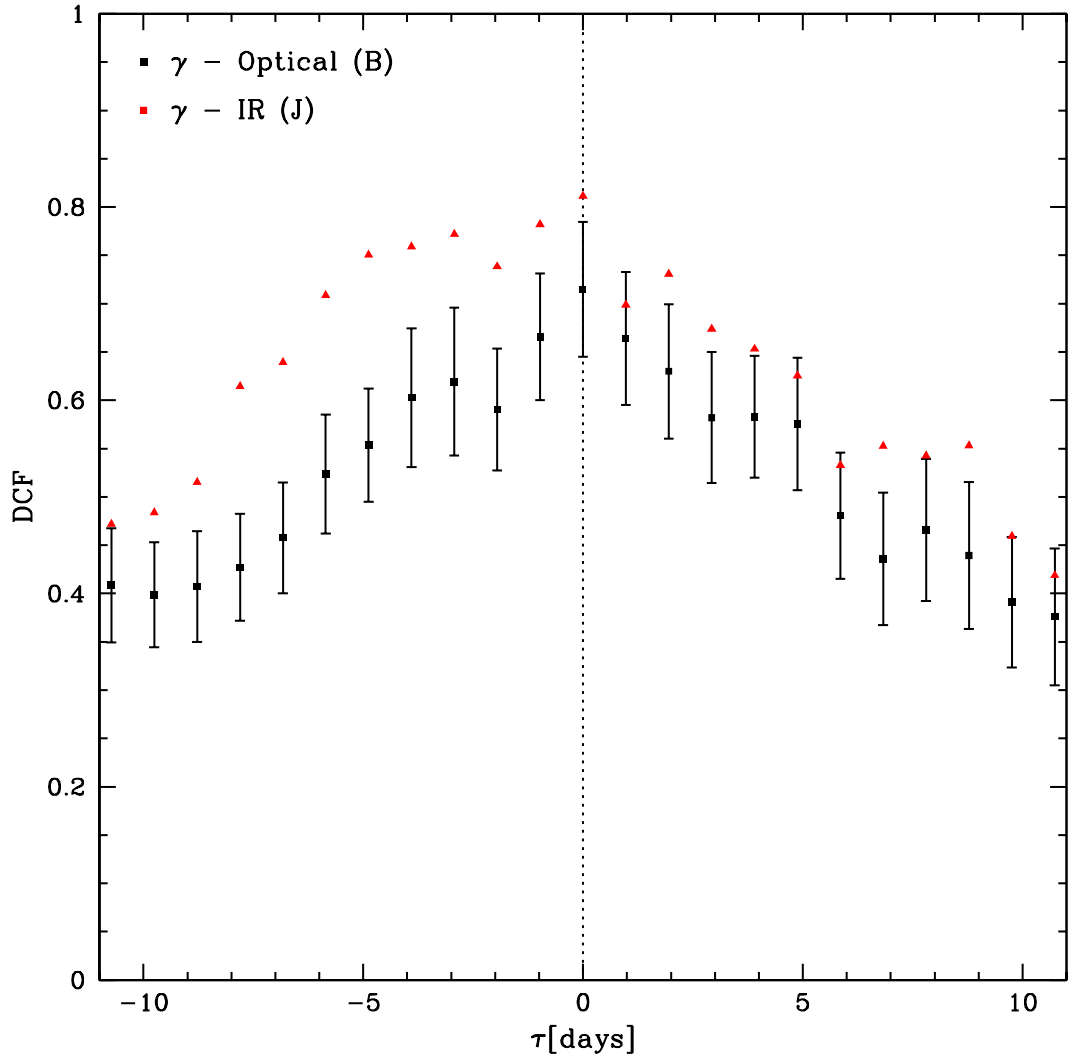} 
\figcaption[fig4]{Discrete correlation function for optical B-band
(black squares) and infrared J-band (red triangles) versus gamma-ray
(0.1--300 GeV) light curves. Error bars on the $\gamma$ - J-band DCF
are comparable to the $\gamma$ - B-band DCF and are omitted for clarity. 
The DCF peaks at zero lag, supporting external Compton models 
in which the variability is due to changes in the spectrum of
relativistic electrons that both radiate the optical/IR 
synchrotron emission and up-scatter soft photons to GeV energies.
The DCF has a hint of a shoulder at $\sim 3-5$~days
(negative lags correspond to the gamma-rays {\em leading} the other
band). This may result from slightly higher electron energies for
the gamma-rays, which would give them shorter radiative timescales. 
\label{fig:dcf}}
\end{center}
\end{figure}

The correlated variability across all observed wavebands except the
X-ray is readily apparent in Figure~\ref{fig:lc}.
All observed bands save the X-ray show two prominent
peaks around JD~2454715 and a short flare near JD~2454740. 
The amplitude is largest in the gamma-rays and J band.
Figure~\ref{fig:dcf} shows the discrete correlation function 
\citep[DCF,][]{edelson88,white94} calculated for 
the gamma-ray (0.1--300 GeV) flux versus light curves in 
the optical B band\footnote{For the B band, we include optical fluxes 
from both the Swift and SMARTS telescopes in order to have complete
coverage over gaps in the individual light curves.},
which has the best temporal coverage, and infrared J-band,
which shows the strongest variations.
The DCF shows a peak correlation amplitude $\sim$0.7 at $\tau = 0$, 
indicating no detectable lag between IR/optical and gamma-ray fluxes. 
Given the sampling, this means any lag is less than or about 1~day.
Similar results were reported by \citet{vercellone09} for
the earlier flare observed with AGILE, though with much lower significance.
The optical versus IR DCF shows even stronger correlation 
(amplitude $\sim$0.8), also with $0 \pm 1$~day lag.

Table~1 shows the fractional root mean square (rms) variability amplitude
\citep{vaughan03} for each band. 
The IR, optical, and UV variability amplitudes decrease
toward shorter wavelengths, suggesting the possible presence 
of steady thermal emission (UV accretion disk emission plus 
Balmer continuum, Fe~{\sc ii}, and \mgii\ in the V and B bands) 
added to the steeper-spectrum jet.
Evidence for `big' and `little' blue bumps was found previously in the
SED of \thesource\ during periods of low emission \citep{raiteri07}. 
The colors of \thesource\ are redder at brighter levels, 
historically \citep{villata06} and in the present data, 
also supporting the presence of thermal
emission beneath the much brighter non-thermal jet. 

The closely correlated IR/gamma-ray variability of \thesource\ 
supports a model in which relativistic electrons in the jet
radiate IR/optical synchrotron photons and inverse Compton scatter 
thermal photons to X- and gamma-ray energies.
The observed gamma-ray flares must be caused by changes 
in the injection luminosity of the higher energy electrons,
rather than variability of the ambient thermal photons, 
since in that case there would be higher amplitude variations 
in the UV than the infrared.
The implication of the short lag time (Fig.~\ref{fig:dcf}) 
is that electrons of similar energy produce IR and gamma-ray emission.

\begin{figure}[]
\begin{center}
\plotone{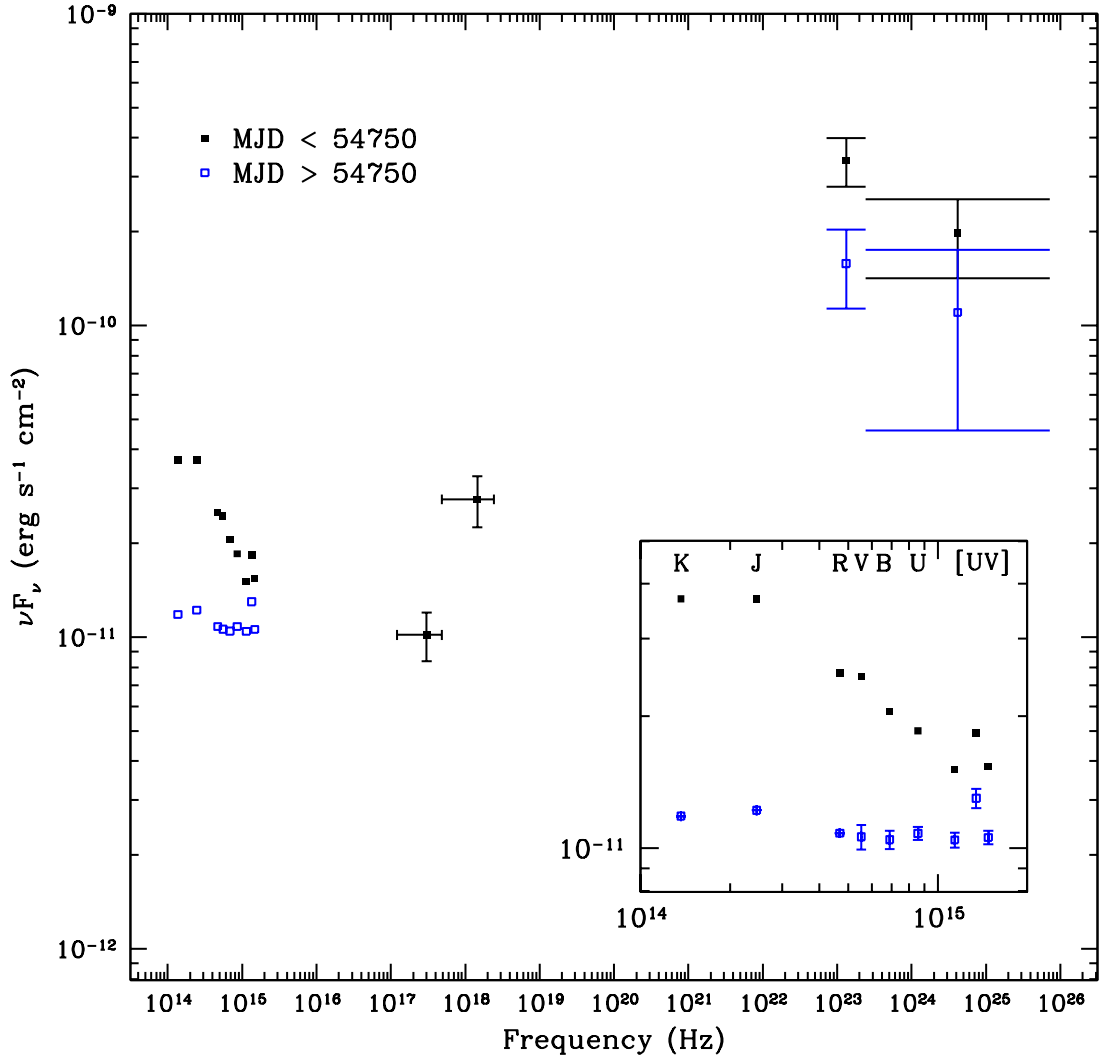} 
\figcaption[fig6]{Spectral energy distributions for high
(JD~2454680--2454750) and low (JD~2454750--2454820) states of
\thesource. Error bars not shown are smaller than
the plotted points. Fermi/LAT fluxes are derived from the
average count rate in the 0.3-1 GeV and 1-300 GeV bands assuming a
power-law spectrum with photon index $\Gamma=2$. The long-wavelength
component in the low state is very flat, not unlike an accretion disk
spectrum, while the variable component is clearly an infrared-bright jet.
\label{fig:sed}}
\end{center}
\end{figure}

Figure~\ref{fig:sed} shows the SED in the high state (JD~2454680--2454750) 
and at the lower final intensity (JD~2454750--2454820). The SED of the
high flux state prior to JD~2454750 shows an optical/IR flux level
similar to that of the May 2007 flare \citep{raiteri08}, intermediate
between the high and low states reported by \citet{raiteri07} (and
references therein), so not surprisingly, the basic model parameters
are similar.  
The optical/UV emission is due to the highest energy electrons 
(Lorentz factors $\sim10^{3-4}$) radiating via synchrotron in a field
of $\sim10$~Gauss, while the gamma-rays come from inverse Compton
scattering on the broad-line photons. The bulk Lorentz factor is
$\Gamma \sim \delta$ $\sim 10-15$  (where $\delta =
[\gamma(1-\beta\cos\theta)]^{-1}$  is the Doppler beaming factor).

The lack of correlation seen in the DCF for 2--10~keV X-rays 
with respect to the other wavebands finds a natural explanation 
in the external Compton scenario, with the X-rays coming from
low-energy electrons ($\gamma\sim$10--100)  
inverse-Compton scattering external UV photons, rather than 
higher energy electrons ($\sim 10^{3-4}$) scattering synchrotron photons.
An SSC component in X-rays would introduce correlation between
X-rays and gamma-rays, which is not seen.
The highest energy electrons (producing the IR/optical and gamma-ray emission) 
vary more rapidly (the radiative timescales are shorter) while the low energy
electrons act as a reservoir and vary more slowly.

More precise SED modeling is needed to determine detailed model parameters,
such as the energy density and location of the thermal photons, 
the location and size of the scattering region, the electron distribution,
the bulk Lorentz factor and jet orientation, etc. This detailed
analysis will be deferred to a later paper. Still, some additional
conclusions can be made. 
The overall stability of source parameters and the correlation 
imply that the emission region is stable on time scales of 
$\sim 1$~month. 
If the electrons are localized in a fast moving knot 
(which might become visible in VLBI maps in a few months), 
it moves a distance $\gamma^2 c \Delta t$, roughly 1-10~pc,
i.e., the jet parameters cannot change dramatically on this scale. 
However, the \citet{sikora94} model for \thesource\ 
can be ruled out as the source of the rapid variations discussed here,
since their assumed source size of $10^19$~cm 
implies $\Delta t \gtrsim 1$~year.
Instead, their model might explain 
a slowly changing, much larger region of the jet.

In conclusion, \thesource\ shows very strong, correlated variability between
the peak of the synchrotron component (at infrared, optical and UV
wavelengths) and the peak of the gamma-ray component.  
No such correlation is seen between X-rays and any other band. 
These results suggest that the variability arises from changes in the
electron luminosity at a compact location in the jet. 
The highly variable infrared through UV emission, particularly in the
brightest state, is dominated by synchrotron emission from a compact
region of high-energy electrons in the jet, with a smaller
contribution from a relatively steady accretion disk. 
The slowly varying low-energy part of the electron spectrum gives rise
to relatively stable X-ray emission via scattering. The gamma-rays vary
in a correlated way because they result from  the same high-energy
electrons up-scattering ambient UV photons.

\acknowledgments
SMARTS observations of LAT-monitored blazars are supported by Fermi
GI grant 011283. CDB, MMB and the SMARTS 1.3m observing queue also
receive support from NSF grant AST-0707627. This research has made use
of the NASA/IPAC Infrared Science Archive, which is operated by the
Jet Propulsion Laboratory, Californina Institute of Technology, under
contract with the National Aeronautics and Space Administration. 

\newpage

\bibliographystyle{apj}
\bibliography{apj-jour,blazar}

\begin{table}[h]
\caption{Fractional Variability Amplitude
}
\centering
\label{table:fvar}
\begin{tabular}{lll}
\hline
Band & F$_{\rm var}$ \\
\hline
K & 0.510 $\pm$ 0.0004 \\
J & 0.603 $\pm$ 0.0001 \\
R & 0.472 $\pm$ 0.001 \\
V & 0.385 $\pm$ 0.001 \\
B & 0.362 $\pm$ 0.001 \\
U & 0.193 $\pm$ 0.002 \\
W1 & 0.165 $\pm$ 0.003 \\
M2 & 0.142 $\pm$ 0.004 \\
W2 & 0.140 $\pm$ 0.004 \\
2-10 keV & 0\tablenotemark{a} \\
0.1-300 GeV & 0.455 $\pm$ 0.015 \\
\noalign{\smallskip}
\end{tabular}
\tablenotetext{a}{The X-ray sample variance was equivalent 
to the mean square error, leading to a value of F$_{\rm var}$ 
consistent with zero.}
\end{table} 

\end{document}